# BUCS: An Engine For Generating Realistic Imaging Data for Deep Galaxy Fields

Rychard J. Bouwens, Garth D. Illingworth, and Daniel K. Magee

*Astronomy Department, University of California, Santa Cruz, CA 95064, Email: bouwens@ucolick.org*

**Abstract.**
Today's deep high resolution multiwavelength surveys contain a wealth of information about galaxies at different epochs. To fully exploit this information, it is useful to be able to produce highly realistic simulations to compare with the observations. Here, we describe one such simulator for producing imaging data for deep galaxy fields. Based upon a pixel-by-pixel modelling of object SEDs and their selection volumes, this simulator (BUCS: Bouwens' Universe Construction Set) allows users to make realistic multicolor simulations of galaxy fields from galaxy samples at all redshifts. Input samples for this simulator range from $z \sim 0$ samples selected from the Sloan Digital Sky Survey to $z \sim 1-6$ samples selected from the Great Observatories Origins Deep Survey. Users can specify the passbands, noise, and PSFs, or equivalently the exposure times on well-known instruments like HST or ground-based telescopes. This simulator provides the community with a real world virtual observatory, useful both at the proposal stage and for making comparisons with observations in hand. The purpose of this presentation is to introduce this tool to the computational community as a whole.

## 1. Introduction

One of the main goals of galaxy evolution studies is to quantify the evolution of galaxy properties across cosmic time. While much progress has been made towards this goal, making sense of this progress in terms of the details has been much more troublesome. The field is literally a mismash of different samples, selection and analysis techniques, and it can often be difficult to interpret the differences that occur between studies. This wouldn't be a large problem if there were a simple means of transforming results from one study to the selection and measurement frame of another. Unfortunately, this cannot typically be done without making additional assumptions since most studies only tabulate derived quantities, like the luminosity or size function.

The principal difficulty is that generic studies of galaxy evolution *do not preserve the total information content* present in the original samples and therefore transformations to other selection or measurement frames can only be approximately done. Real galaxy samples consist of disparate sets of individual galaxies, each with their own pixel-by-pixel multiwavelength morphologies, volume den-sities, and interobject spacing. Choosing to represent these samples exclusively with select numbers of derived quantities (e.g., the LF, size functions) results in a loss of information and can compromise the usefulness of these samples to other scientists.

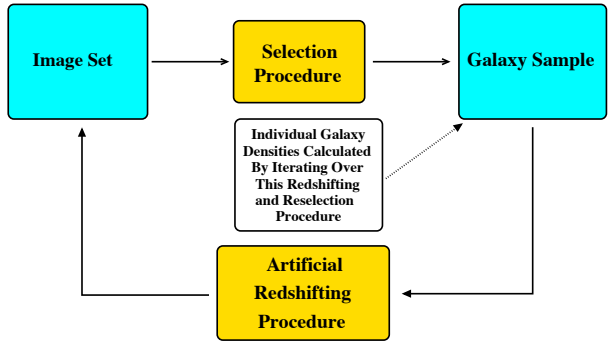

Figure 1. Schematic illustrating the general flow of the BUCS software library. The software automates the conversion of image sets (e.g., deep galaxy fields like the HUDF) into galaxy samples. Galaxy samples can then in turn be laid down over a specific redshift interval and thus used to generate image sets. Volume densities of individual galaxies can be determined by artificially redshifting each galaxy template to all redshifts and reselecting it using the selection criteria of the sample. Image sets are essentially what they sound like: coaligned sets of images in different filters, with PSFs, zero points, and noise properties. Samples can be anything from a standard $B$-dropout sample selected using two colors from the HUDF to a $z \sim 0.3$ spectroscopic sample selected from GOODS.

At the same time, the shear quantity of multiwavelength data available on samples of high-redshift galaxies continues to increase. Eight years ago, essentially the only field with multiwavelength, high-resolution data was the HDF North. Now, with the availability of the NICMOS and ACS on HST, Chandra, Spitzer, GALEX, and $8-10$-m ground-based telescopes, the data volume has increased at least 100 fold. While there have also been noticeable advances in computing speed, processing these data by hand to produce galaxy samples continues to be more and more tedious and time consuming. Telescopes like the LSST can only make these matters worse.

Foreseeing these trends, several years ago we set about to develop some software to automate the process. This software eventually became known as BUCS (Bouwens' Universe Construction Set) and was centered around the concept of a galaxy sample and the extraction of samples from deep galaxy images. The reason we decided to focus on galaxy samples is that these samples really act as our stepping stones for evaluating evolution across cosmic time, and it seemed obvious that procedures would be needed to quickly and easily convert observational data into these more scientifically meaningful units. It was also evident that such procedures would also provide a convenient way of preserving the total information content in such samples (i.e., multiwavelength pixel-by-pixel fluxes on all individual sample objects) so that they could be readily disseminated for comparison with newer samples being acquired.

## 2. Overview

The BUCS software suite is centered around the abstraction of a galaxy sample. Galaxy samples are defined by an image set, selection criteria, and a procedure for estimating redshifts. With this software, deep galaxy image sets can be seemlessly converted into galaxy samples. Conversely, galaxy samples can be used to simulate sections of the universe and hence produce image sets again. A simple illustration of the general flow is provided in Figure 1.

The software is designed to manipulate samples, image sets, and individual galaxies in much the same way that other software libraries manipulate the abstract objects which are the focus of those libraries (e.g., as cfitsio allows one to manipulate fits files). It should therefore be no surprise that basic calculations like the determination of sample selection functions or the projection of one sample onto another are automated in the BUCS software suite (Figure 1). This latter function naturally allows for tests of the null-evolution hypothesis and hence some measure of the actual evolution.

After years of development, this software is now quite mature. One testament to this is the large number of refereed publications which have already made heavy use of it (Bouwens et al. 1998a,b; Bouwens et al. 2003a,b; Bouwens et al. 2004a,b,c). Dissemination of the software and some much needed documentation should be available in the not too distant future.

## 3. WBUCS: A Web Simulator for Deep Galaxy Fields

By automating the procedure for quantifying the volume density of individual galaxies in different samples and modelling the pixel-by-pixel SEDs of each object, it is straightforward to use this information to simulate deep galaxy fields and to make this available to the community at large. In our implementation of this (see P.91 from this volume), we utilize galaxy samples at all redshifts, ranging from the Sloan Digital Sky Survey at $z \sim 0$ (Blanton et al. 2004) to a low-redshift GOODS sample at $z \sim 0.2 - 0.4$ (Bouwens et al. 2004e) to a wide variety of different dropout samples from $z \sim 2$ to $z \sim 6$ (Bouwens et al. 2004d). This service should be useful both at the proposal stage and for comparing with existing data.

**Acknowledgments.** We are grateful to Narciso Benítez, Emmanuel Bertin, Tom Broadhurst, Fred Courbin, Michele de la Pena, Harry Ferguson, Marijn Franx, Andy Fruchter, Jon McCann, and Gerhardt Meurer for input important for the development of this software. We also gladly acknowledge fruitful collaborations with other members of the ACS GTO team and support from NASA grant NAG5-7697.

## References

Blanton, M.R., et al. 2004, AJ, submitted, astro-ph/0410166.

Figure 2. Users request simulations of deep galaxy fields through simple forms like the one shown at the top of this figure. The output is composed of simple fits images of the field, as viewed through different passbands (shown at the bottom of this figure). Once the request has been received by our servers, it is placed in the queue and an e-mail is sent out summarizing the request. After the job is completed, the simulated images are placed at an ftp site and the user is notified as to the location where they can download them.

Bouwens, R.J., Broadhurst, T.J., & Silk, J. 1998a, ApJ, 506, 557.
Bouwens, R.J., Broadhurst, T.J., & Silk, J. 1998b, ApJ, 506, 579.
Bouwens, R.J., et al. 2003a, ApJ, 595, 589.
Bouwens, R.J., Broadhurst, T.J., & Illingworth, G.D. 2003b, ApJ, 593, 640.
Bouwens, R.J., et al. 2004a, ApJ, 606, L25.
Bouwens, R.J., et al. 2004b, ApJ, 611, L1.
Bouwens, R.J., et al. 2004c, ApJ, 616, L79.
Bouwens, R.J., et al. 2004d, ApJ, submitted.
Bouwens, R.J., et al. 2004e, ApJ, in preparation.
Magee, D.K., Bouwens, R.J., & Illingworth, G.D. 2005, will be given at a later time, [P.91].